# Microwave Frequency Fiber Interferometry in Submarine Deployed Telecommunication Cables

A. Bogris[1], C. Simos[2], I. Simos[3], Y. Wang[4], A. Fichtner[4], S. Deligiannidis[1], N. S. Melis[5], C. Mesaritakis[6]

*1. Department of Informatics and Computer Engineering, University of West Attica, Aghiou Spiridonos, 12243, Egaleo, Greece, 2. Dept. of Physics, University of Thessaly, 35100, Lamia, Greece 3. Department of Electrical and Electronics Engineering, University of West Attica, 12244 Egaleo, Greece, 4. Department of Earth and Planetary Sciences, ETH Zurich, Zurich, Switzerland, 5. National Observatory of Athens, Institute of Geodynamics, Athens, Greece, 6. Department of Biomedical Engineering, University of West Attica, Aghiou Spiridonos, 12243, Egaleo, Greece*
*e-mail address: abogris@uniwa.gr*

**Abstract:** We operated a microwave frequency fiber interferometer in a telecommunication cable in the Ionian Sea, Greece, for two months. The capability of detecting undersea micro earthquakes (magnitude~1.5), tides and ocean waves is reported. © 2025 The Author(s)

## 1. Introduction

The advent of fiber-optic sensing in commercially deployed cables gains traction as it may transform fiber infrastructure into a large sensor network for environmental monitoring, network monitoring and resilience, and for critical infrastructures security. The dominant fiber-optic sensing technique is distributed acoustic sensing (DAS), which offers meter-scale resolution and nanostrain sensitivity [1]. However, its use becomes challenging in long cables (>50 km) usually deployed in submarine environments. Interferometric or polarization monitoring techniques, although less capable in offering localization, are compatible with long-haul transmission and constitute significant techniques for monitoring the deep ocean, where other sensors are difficult to deploy [2-4]. Ultra-stable optical interferometry [3] is a highly sensitive technique for strain measurements for geophysical monitoring. However, it cannot be widely deployed because its stability relies on Hz-linewidth laser sources, which are costly and not easy to fabricate in mass production. Polarization sensing is practical, as it uses data captured by commercially deployed transceivers, but its sensitivity is lower than that of phase detection [4]. Moreover, taking into account that coherent receivers are developed to track multi-GHz frequency bandwidths, new polarization sensing receivers must be implemented with focus on the Hz-level frequency window of mechanical vibrations to offer quasi-real time sensing and event detection. Bogris et al. [5] presented the microwave counterpart of optical interferometry, the microwave frequency fiber interferometer (MFFI), as a potentially more practical integrated sensing technique for sensitive strain measurements across deployed cables. The stability of MFFI relies on the mature technology of microwave oscillators with sub-Hz frequency stability, which makes robust interferometric measurements possible even in long transoceanic links. Although the principle of operation and its ability to detect earthquakes with magnitudes as low as 3 has been demonstrated in terrestrial links, MFFI has never been utilized in submarine links where other effects such as tidal waves, storm surges or ocean waves are present. In this paper we report what is to our knowledge the first deployment of MFFI in a submarine environment around the island of Cephalonia, Greece, which is characterized by elevated earthquake activity. We successfully detected micro earthquakes (magnitude ~1.5), tidal waves and ocean waves. The results are in remarkable agreement with data recorded by the accelerometers of the National Observatory of Athens (NOA) and DAS provided by a Silixa iDAS interrogator. These experiments demonstrate that MFFI is a very promising low-cost and portable candidate for high-sensitivity geophysical monitoring and early warning for natural hazards in the deep ocean.

## 2. Field trial and experimental setup description

Our experiment took place on the island of Cephalonia, near the town of Fiscardo (fig. 1a). The fiber link was provided by the Hellenic Telecommunications Organisation S.A. (OTE) and is 15.6 km long. It connects Cephalonia with Ithaca, and consists of 7.2 km of terrestrial cable in Cephalonia, 7.1 km of submarine cable in the sea separating the two islands and finally 1.3 km of terrestrial cable in Ithaca (fig. 1a). The MFFI system was installed at Antipata together with a Silixa iDAS on 24 April 2024 and both operated for two months. To have a unique fiber path for each interrogator, we used two separate fibers and applied a loop-back on the side of Ithaca for MFFI path. Compared to what has been demonstrated in [5], the MFFI has been improved in the electronics part with a higher resolution, broader bandwidth, lower-noise analog-to-digital converter (ADC, 800 samples/s, 16 bit resolution), which substantially improved the sensitivity of the measurements at small phase shifts. The new ADC offers a resolution on the order of 0.02 mrad, almost 30 times better than that of [5]. The location of the fiber is close to one of the most seismically active areas of Greece, near the Cephalonia transform fault (fig. 1b). During the experiment several clusters of earthquakes with small magnitudes (1.5 - 3.0) could be observed (fig 1b). This gave the opportunity to investigate

MFFI signals compared to signals recorded by seismic stations operated by NOA, as part of the Unified Hellenic National Seismic Network (fig 1a). We compare records especially with Fiskardo station (FSK.HP) recordings, sited next to the fiber (fig. 1a). Besides earthquake detection, the submarine part of the cable gave the opportunity to verify that MFFI is capable of recording tidal and ocean waves, as will be shown below.

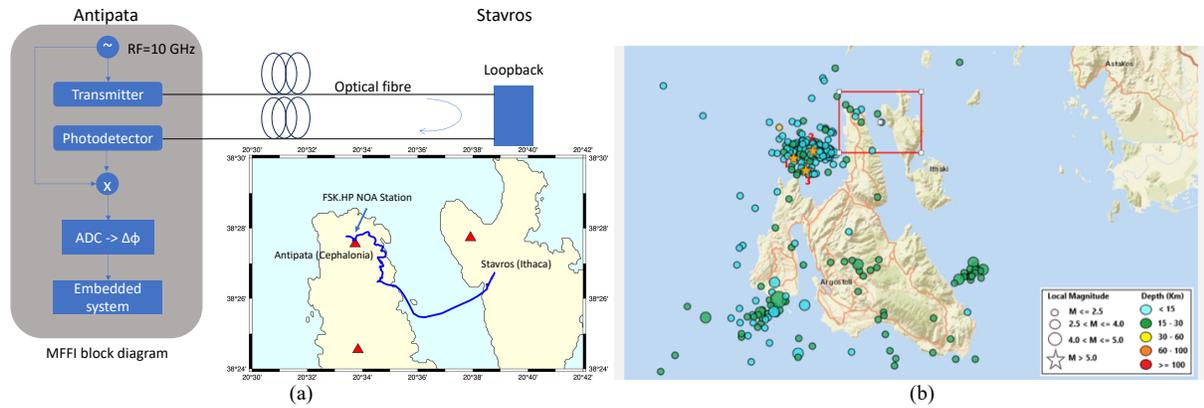

Fig. 1: a) MFFI building blocks and the map showing the cable route in land and sea areas. The NOA accelerometer station at Fiskardo (FSK.HP) is also depicted in the map (red triangle), south close to the fiber in blue. b) Local seismicity during the experiment located by NOA (stars with numbers denote the 3 events in fig. 2 and red box the experiment area).

## 3. Results

During the experiment, local seismicity included more than 200 events recorded by NOA's stations. The majority of these form a cluster in the sea west of Antipata (fig. 1b). Although a complete analysis of the data is still in progress, we already identified >110 earthquakes with magnitude between 1.5 to 3.0. The sensitivity was validated by comparing our system to the recordings at NOA station FSK.HP. Fig. 2 demonstrates the excellent agreement between the MFFI data and the vertical and horizontal components of FSK.HP, including P and S wave arrivals for both events.

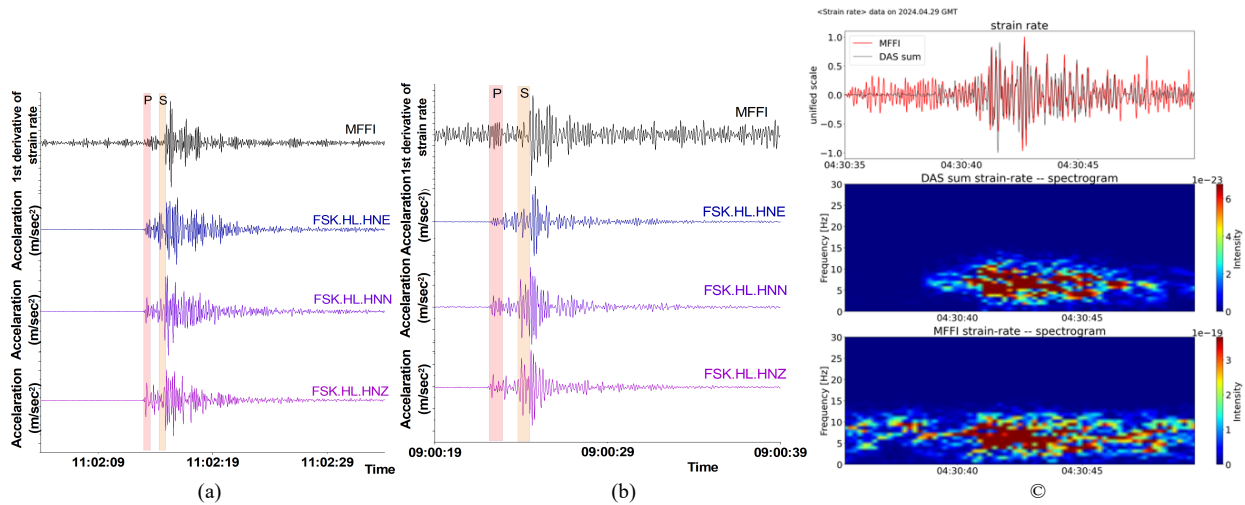

Fig. 2: Comparison of time series from MFFI and FSK.HP for earthquakes 1 and 2 in fig. 1b. a) 25 April 2024, 11:02:11 UTC, magnitude 2.5, b) 9 May 2024, 09:00:21 UTC, magnitude 1.6). P and S arrival times are marked and clearly visible in MFFI data. c) Comparison of MFFI strain rate and DAS average strain rate (earthquake 3: 29 April 2024, 04:30:35 UTC, magnitude 2.1).

It is evident that MFFI is capable of measuring micro-seismic activity (seen events with magnitude 1.6). These events were observed using simple signal processing, such as low-pass filtering. More sophisticated processing, including de-noising, is expected to reveal even lower magnitude events. The agreement between DAS average strain rate and MFFI strain rate results is remarkable (fig. 2c, earthquake 3). Even the coda waves are well synchronized. Regarding the spectrograms in the same figure, we clearly see that a DAS channel located 10.2 km from Antipata, located under water, has clear similarity with an MFFI spectrogram during the earthquake. The MFFI spectrogram is also affected by noise sources in the terrestrial part.

We also evaluated MFFI as a potential interrogator of oceanic processes. We first analyzed the data for a long period (1 to 15 June 2024) to identify tidal waves. Fig. 3a shows the spectrum of strain for this period, which contains two clear peaks, one at 24 hours (related to temperature changes mostly affecting the terrestrial part of the cable) and one at 12 hours (relate to the tidal waves affecting the submarine part of the cable). Based on Meteo.gr, the standard Greek weather forecasting website and service operated by NOA, we also identified days with different wind speed and with north-west orientation that could affect the sea area and see if they could influence the power-spectral density at frequencies related to sea waves (0.04 - 0.1 Hz). As an example, we chose May 4th, which was the windiest day in May (14:20 wind speed 59.5 km/h) and May 8th, which was one of the calmest days of the month (average speed 1.7 km/h, highest speed 17.7 km/h at 00:40). In Fig. 3b we depict the MFFI strain rate frequency response within an hour including the maximum wind speed per day, that is 14:00-15:00 for May 4th and 00:00-01:00 for May 8th. The spectra clearly show 5 dB higher power-spectral density for May 4th data in a frequency range related to sea waves. Similar results have been systematically obtained for other days with different wind speeds, confirming that MFFI is also capable of detecting sea waves.

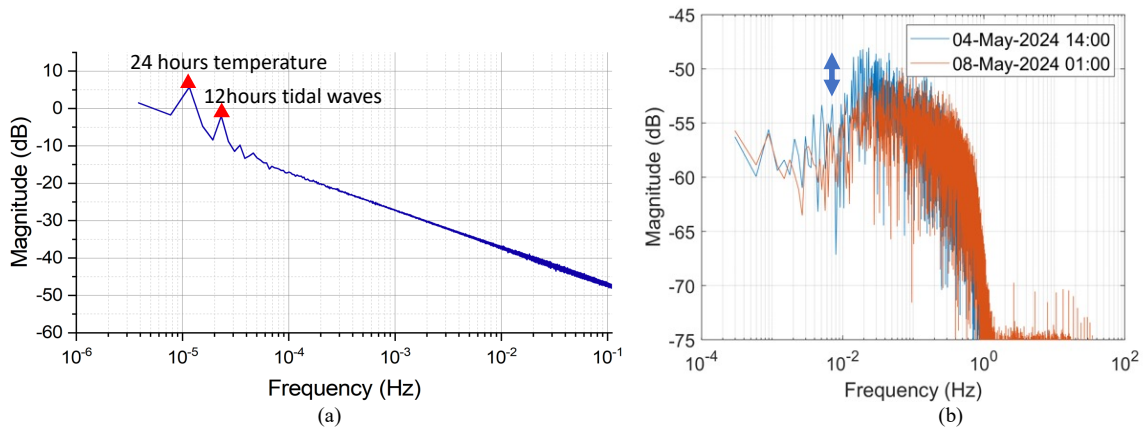

Fig. 3: MFFI long term strain measurement and their frequency response. a) Tidal waves and temperature changes. b) Sea waves and their relation to wind speed.

## 4. Conclusions

We demonstrated the interrogation of submarine cables by MFFI in a field experiment from 24 April 2024 to 24 June 2024 at Cephalonia Island in the Ionian Sea. The experimentation with MFFI revealed its efficacy in detecting micro-seismic activity with magnitudes as low as 1.5 and ocean processes such as sea and tidal waves. Taking into account that MFFI interrogators are simple to implement, this work demonstrates that it is a promising solution for low-cost, robust and high-sensitivity geophysical monitoring in the oceans.

We would like to thank I. Chochliouros, C. Avdoulos, A. Apostolatos and D. Alissandratos from OTE for providing access to the fiber and assisting in MFFI and DAS installation at Antipata. This work has been partially funded by Horizon Europe Project QPIC1550 under Grant Agreement 101135785 and ECSTATIC under Grant Agreement 101189595.